\documentclass{article}

\usepackage{PRIMEarxiv}

\usepackage[utf8]{inputenc} 
\usepackage[T1]{fontenc}    
\usepackage{hyperref}       
\usepackage{url}            
\usepackage{booktabs}       
\usepackage{amsfonts}       
\usepackage{nicefrac}       
\usepackage{microtype}      
\usepackage{lipsum}
\usepackage{fancyhdr}       
\usepackage{graphicx}       
\graphicspath{{media/}}     
\usepackage{amssymb}
\usepackage{amsmath}
\usepackage{listings}
\usepackage{cuted}
\usepackage{algorithm}
\usepackage{algorithmicx}
\usepackage{algpseudocode}
\title{Double-Precision Floating-Point Data Visualizations Using Vulkan API}

\author{
  Nezihe Sözen \\
  Siemens Turkiye \\
  Istanbul\\
  \texttt{nezihe.sozen@siemens.com} \\
}

\begin{document}
\maketitle

\begin{abstract}
Proper representation of data in graphical visualizations becomes challenging when high accuracy in data types is required, especially in those situations where the difference between double-precision floating-point and single-precision floating-point values makes a significant difference. Some of the limitations of using single-precision over double-precision include lesser accuracy, which accumulates errors over time, and poor modeling of large or small numbers. In such scenarios, emulated double precision is often used as a solution. The proposed methodology uses a modern GPU pipeline and graphics library API specifications to use native double precision. In this research, the approach is implemented using the Vulkan API, C++, and GLSL. Experimental evaluation with a series of experiments on 2D and 3D point datasets is proposed to indicate the effectiveness of the approach. This evaluates performance comparisons between native double-precision implementations against their emulated double-precision approaches with respect to rendering performance and accuracy. This study provides insight into the benefits of using native double-precision in graphical applications, denoting limitations and problems with emulated double-precision usages. These results improve the general understanding of the precision involved in graphical visualizations and assist developers in making decisions about which precision methods to use during their applications.
\end{abstract}


\section{Introduction}
Nowadays, computer graphics tools play a significant role in high-precision data modeling and display for several themes within science and engineering. They make an impact on decision-making and consequent results in very diverse domains, from scientific research to gaming. High-precision molecular dynamic simulation \cite{jia2020pushing}, an attempt at film-game integration \cite{lou2023lost}, next-generation AI-based games \cite{zhao2022deep}, the open-source dynamic simulation framework YADE \cite{kozicki2022implementation}, three-dimensional reconstruction using \cite{cui2021high}, and ray tracing of CAD-based simulation \cite{shriwise2022hardware} are some of the studies which underpin the need.

In graphical models, precision and accuracy are vital and are mostly used in scientific computing, medical imaging, and engineering simulations. These models guarantee to convey complex and comprehensive information precisely and accurately. The ancillary problem here is obtaining correct data types, mainly distinguishing double-precision from single-precision floating-point types, which turn out to be very influential. This kind of technology is, however, vision-limited to some level of accuracy, hence less valuable if even higher levels of precision are called for. It is, hence, important to pick an alternative that offers precision and, importantly, the level of accuracy required by the application or the project.

Processing and visualizing double-precision data present several challenges, such as computational intensity \cite{henry2019leveraging}, hardware and software support \cite{alidori2020hardware} \cite{said2021fpu}, and rounding errors \cite{jezequel2020can} \cite{croci2022stochastic}, mainly due to the complexity and precision requirements of this type of data.

The choice of API in high-accuracy applications is based on each API's appropriateness. With changing application requirements, the suitability of each API changes: OpenGL \cite{OpenGL}, Vulkan \cite{Vulkan}, DirectX \cite{DirectX}, OpenCL \cite{OpenCL}, and CUDA \cite{CUDA} . Vulkan is a much newer API compared to OpenGL and gives a developer much more acute control over GPU processes, significantly improving rendering performance and speed. Direct3D is the most frequently used graphics API for game and multimedia applications that run on Windows systems. OpenCL is a portable programming language that can be run on various devices, hence the reason why OpenCL applications are highly portable across a wide variation of hardware like a GPU and CPU.

In the case of high-precision visual representation using Vulkan API and GLSL \cite{CoreLanguageGLSL}, complicated technical details shall be addressed in the analysis of double-precision and single-precision floating-point data. One major reason could be the accuracy differences that, in turn, compromise both performance and memory consumption. In old generations of GPU hardware, a representation of double-precision data often emulated single-precision floating-point numbers. Modern GPUs have, however, come with support for double-precision data in their design. This research is based on recent versions of GLSL supporting both single and double precision. The evaluations taken into consideration for this research are to frame the differences between single-precision and double-precision data representations. In this paper, the approaches are compared with respect to visualization performance, where a rendering operation differs in both. This comparison will provide important information about which graphics API is more suitable for applications requiring high precision and will provide guidance for current applications in the field of graphics programming. Additionally, current challenges in processing and visualizing double-precision floating-point data and methods to overcome these challenges will be discussed.

In view of the requirement of accuracy in graphics visualizations, primarily due to real-world applications in science and engineering, this research is done to find out how double-precision floating-point data visualizations realize performance gains using the Vulkan API. The specific research questions that guided this study were as follows:
\begin{itemize}
    \item RQ 1. How does the performance of native double-precision floating-point implementations compare to emulated double-precision implementations in Vulkan for rendering 2D and 3D points datasets? 
    \item RQ 2. How does the scalability of double-precision floating-point data visualization in Vulkan API hold up with increasing dataset sizes?
\end{itemize}

\section{Related Work}
This section reviews previous studies on the use of double- and single-precision floating-point values in graphics processing units (GPUs) and discusses the findings, similarities, and differences of these studies in the context of current research. Various floating-point approaches have been developed using graphics APIs and game engines. However, this section examines closely related approaches like emulated double-precision, hardware and software-based double-precision, and extended precision.

Da Graça \& Defour \cite{DBLP:journals/corr/abs-cs-0603115} demonstrated a 44-bit solution emulating floating-point formats and corresponding operations, which increased the precision for applications that need more than the single precision, complemented with detailed performance and accuracy results. This implementation enabled straightforward and efficient operations for adding, multiplying, and storing floating-point numbers. It is shown that the research in compensated algorithms with float-float representation runs more efficiently for comparable accuracy, and adapting these algorithms to the GPU constitutes a significant part of future research.

Thall \cite{10.1145/1179622.1179682} introduced 'doublefloats' for extended precision representation of floating point numbers for GPU computation. This representation, while improving accuracy without performance degradation on a GPU, comes at the cost of resources by exploiting the inherent parallelism in it. Doublefloats are constructed from unevaluated sums of 32-bit floats and deliver a precision of 48 significant bits. This approach—crucial for very high precision applications—necessarily restricted the use of GPU hardware. This shows, with the help of the Mandelbrot Set Explorer, both the utility of doublefloats and some potential applications in simulation, scientiﬁc computing, and image analysis. 

In the OpenSpace \cite{8370192} study,  the limitations of floating-point numbers in computers and the various methods developed to ensure accurate representation of large-scale astronomical data are discussed. With the method it offers, OpenSpace manages to solve precision problems by using Dynamic Scene Graph and rendering objects at the correct distances relative to the camera in cases where single-precision floating-point numbers are insufficient. This method enables accurate and efficient visualization of large-scale astronomical data and minimizes floating-point precision problems.

Dally et al. \cite{dally2021evolution} review the progress of GPUs from special-purpose hardware for 3D graphics to powerful programmable processors applied in HPC and deep learning. It reflects all significant steps of this development, including the creation of CUDA, the introduction of double-precision floating-point arithmetic, and other innovations like Tensor Cores that have increased the performance and flexibility of contemporary GPUs manyfold. These results depict that in the future, a GPU will keep evolving to provide high performance and support many applications.

Kaufmann et al. \cite{kaufmann2021accurate} addresses the challenges and limitations inherent in real-time physics simulations in large-scale environments, primarily due to the imprecision of single-precision floating-point calculations. The authors solve a limitation where traditional physics engines still rely on single-precision floating-point numbers by proposing a system in which the subdivision of the simulation world into independent sectors takes place, and these sectors are allocated dynamically. It drastically reduces the occurrence of precision errors through cloning at sector boundaries, ensuring very consistent and accurate interaction across these sectors. It has hugely improved the precision and efficiency of real-time physics simulation in large-scale virtual environments by dividing the world into independently simulated, dynamically allocated sectors and using a cloning system to maintain accurate interactions at sector boundaries.

The progress report \cite{Godot} on the Godot Engine reviews challenges to render big worlds in games with single-precision floating-point numbers that lead to precision errors and jerky motions. The report explores solutions such as using double-precision in calculations but handling its impracticability again due to the limits of GPU. The final solution considered is emulating double precision by using two single precision floats for specific matrix calculations, where they preserve the accuracy and precision without much penalty in performance. 
\section{Background and Motivations}
In this section, a number of the fundamental technologies and methodologies of computer graphics will be introduced, including the pipelines of modern Graphics Processing Units (GPUs), the functionality of key graphics APIs, for instance, OpenGL and Vulkan, the use of the Graphics Shader Language (GLSL), the key differences between single and double precision operations, and finally, common issues with single precision floating point computations.

\subsection{GPU Pipeline}
The Graphics Processing Unit pipeline has become a critical component in modern computing, having extensive applications in graphics rendering and general-purpose computing. This paper explores several aspects of GPU pipeline advancements and applications as described in recent literature.

All pipeline stages concerning rasterization, pixel processing, and abstract geometry processing are implemented on the GPU. Internally, these are divided into a range of hardware stages with differing levels of programmability or configurability. The API provides an access method for the programmer to the logical model of a GPU; the actual implementation of this conceptual process in hardware is left to the manufacturer. The pipeline of the GPU processes graphical input through many phases, from the original vertices to the final pixel rendering, with differing degrees of programmability. The fully programmable vertex shader stage is responsible for perspective projection, lighting, model space to view space transformations and vertex shading. Another programmable stage is Geometry Shader, which deals with entire primitives to generate or to modify them and construct complex effects, including particle systems and shadow volumes. 

\begin{figure}
    \centering
    \includegraphics[width=1.0\linewidth]{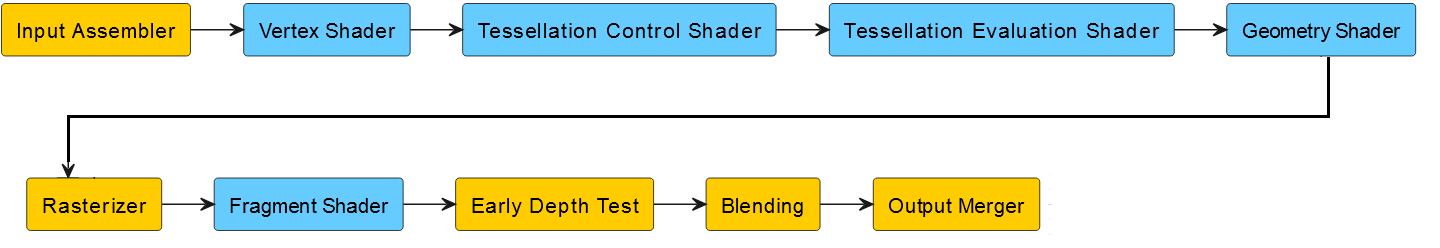}
    \caption{Beginning with the input assembler, which takes the vertex data to assemble vertices into primitives, the pipeline is followed by a vertex shader for geometric transformations. Then, there is a tessellation control shader that performs the division of the surface, followed by the tessellation evaluation shader, refining the vertices. Subsequently, there is a geometry shader for generating or modifying geometry. Thereafter, it proceeds to the rasterizer, which projects 3D primitives onto the 2D screen. Next up is a fragment shader to compute pixel attributes, an early depth test optimization by discarding occluded fragments, the blending stage, which combines fragment colors, among other things, for transparent effects, and an output merger that finally writes the image in the frame buffer for display.}
    \label{fig:enter-label}
\end{figure}

Stream output reuses processed data and is thus useful for effects like hair rendering. Fixed-function stages are triangle preparation and traversal, where triangles are prepared and rasterized into pieces; screen mapping, which translates the vertices from clip space into screen space; and clipping, which cuts triangles beyond the viewing frustum. There may be removal of occluded pieces based on the early z-test step, which is varied across GPUs for increased efficiency. The programmable pixel shader processes every fragment to provide texture and color effects. The final step, raster ops or blending, is where colors are combined, and other pixel tests, like depth and alpha testing, are managed. This pipeline consists of both fixed and programmable steps; both are required to efficiently render complex images. \cite{gregory2018game} \cite{akenine2019real}. 

\subsection{Vulkan API}
Vulkan is the next-generation, efficient, and cross-platform graphics and compute API for enabling access to modern GPUs in today's devices—PCs, consoles, mobile phones, and even embedded platforms. The Vulkan API has been designed to give much more direct control over the GPU, thus allowing finer-grained optimizations and efficient usage of the GPU. Vulkan significantly reduces driver overhead compared to older graphics APIs. Such overhead may yield great performance, particularly in CPU-bound applications. It also designs the API to be more predictable and with fewer errors, clear performance benefits from keeping the GPU busier, producing fewer bottlenecks than those caused by the CPU \cite{bailey2023vulkan}. Vulkan is characterized by its verbosity and fragility, but it provides enhanced control, a streamlined threading architecture, and superior performance. It provides functionality for transport, computation, and graphics and may be chosen as an option \cite{sellers2016vulkan}. The Vulkan Specification mandates a host environment with runtime support for 8-16, 32, and 64-bit signed and unsigned twos-complement integers, 32- and 64-bit floating-point types satisfying range and precision constraints, and ensuring their representation and endianness match those on every supported physical device.

\subsection{GLSL}
GLSL, often known as the OpenGL Shading Language, has a crucial function in contemporary computer graphics by enabling programmatic control over the graphical processing pipeline. This programming language provides a whole set of tools with which developers can create very flexible shaders, improving the ability to develop complex, dynamic graphical effects vastly in any real-time application.The GLSL has become central to a modern graphics programmer due to its wide application in different domains, from game development and virtual reality to scientific visualization. Recent developments in this area include the integration of GLSL with all major graphics APIs and its application in parallel computing cases. Unlocking the doors of new frontiers in graphical rendering and visualization is possible with GLSL. Thereon, the further development of GLSL and the expansion of the practical spectrum of its implementation essentially volatilely characterized the area of computer graphics when new challenges and prospects succeed one another. \cite{10.1145/2077434.2077446} \cite{bailey2009graphics}

\subsection{An Explanation of the Distinction Between Single and Double Precision} 
IEEE 754 floating-point format represents all the standards; it includes 32-bit single precision, 64-bit double precision, and an extended precision format. Each format includes a sign bit, an exponent section, and the mantissa part (fraction) \cite{8739150}. 

The single-precision floating-point format defined in IEEE Std 754-2019 utilizes a 32-bit (4-byte) structure. This format consists of a 1-bit sign bit, telling whether a number is positive or negative, an 8-bit exponent defining the scale of the number adjusted by a predefined "bias" value, and a 23-bit fraction representing the significant or mantissa part of the number. The single-precision format provides an accuracy of about 7 decimal digits while it covers a very wide range of values. This would normally be used in cases where speed is very essential and very fine precision is not required, for example, in the processing of graphics or audio \cite{Johns} \cite{Sanglard}.

\begin{figure} [H]
    \centering
    \includegraphics[width=1.0\linewidth]{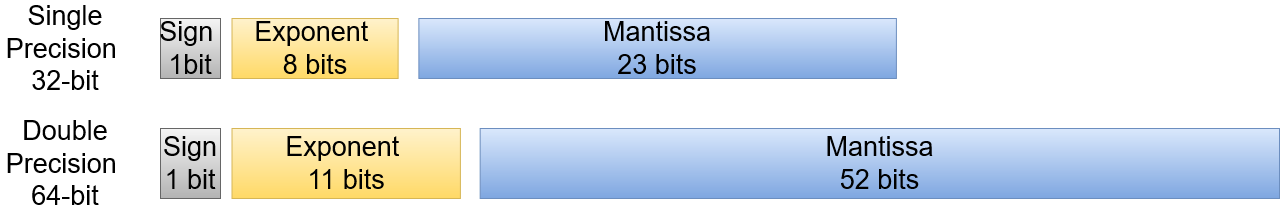}
    \caption{This image is a bit-layout of single and double-precision floating-point numbers, as represented in accordance with the IEEE 754 standard. Single precision number would be 32 bits long. Bits needed for this: 1 bit for the sign, 8 bits for exponent, and 23 bits for mantissa. Double precision number: it is 64 bits; 1 bit for the sign, 11 bits for exponent, and 52 bits for mantissa. It has a wider range and is more accurate in representing floating-point numbers.}
    \label{fig:enter-label}
\end{figure}

The format, otherwise known as double-precision floating point, is 64-bit (8-byte) in size. Much like the single precision, the format contains a 1-bit sign bit but reserves 11 bits for the exponent and 52 bits for the fraction. These give double precision a much greater range and much higher precision. This format provides an accuracy of about 15–16 decimal digits and is preferably used where high accuracy is required, like in scientific computations and precision engineering tasks.

The IEEE 754 standard standardizes the way of representation and processing of floating-point numbers within a computer system. This creates consistency and reliability for numerical computations. The standard provides for the accuracy of numerical operations across different systems and by different languages. This is very important in scenarios where different applications and cross-platform are required.

\subsection{Problems with Single-precision floating point}
The prevailing solution to obtain high precision in graphical visualizations is using double-precision floating-point values. Doubles give a maximum of about 15 to 16 decimal digits of precision and, at the same time, offer far greater range than single-precision, floating-point values. This rise in precision and range drastically reduces errors due to rounding, positioning, accumulation, overflow, underflow, and limitation \cite{DoublePrecisionCoordinates}. In order to understand more easily the problems caused by single-precision floating point, the Mandelbrot Set \cite{mandelbrot2004fractals} formula has been used and rendered. A Mandelbrot set is a mathematical set that repeats in a certain way in the complex plane. 

\begin{table}
\centering
\begin{tabular}{ccc}
\toprule
\includegraphics[width=0.33\textwidth]{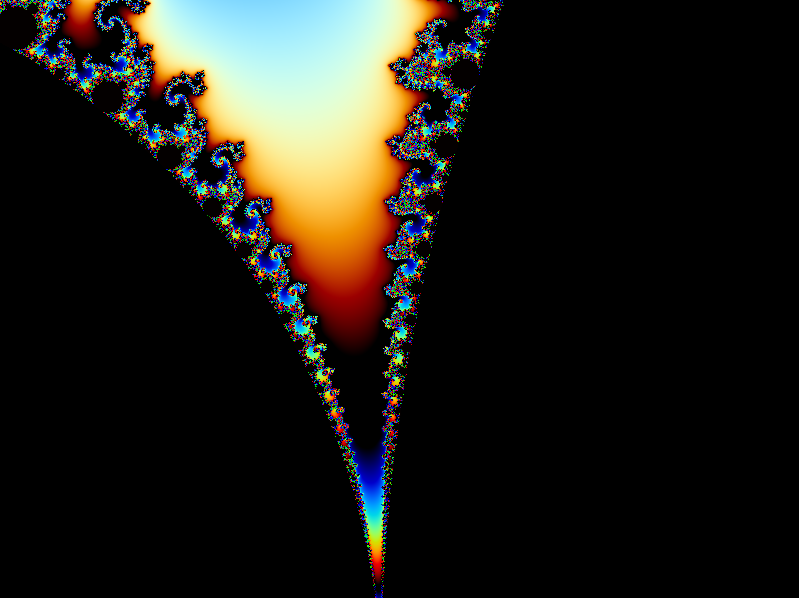} &
\includegraphics[width=0.33\textwidth]{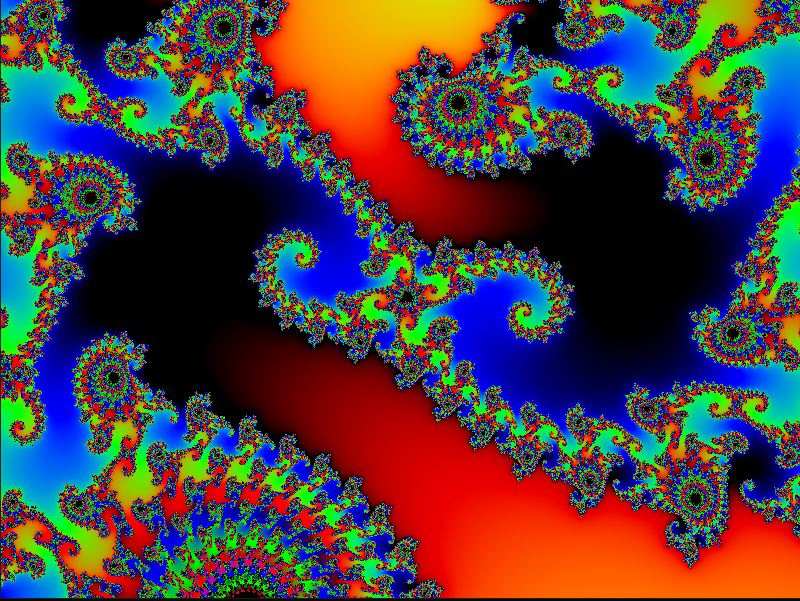} &
\includegraphics[width=0.33\textwidth]{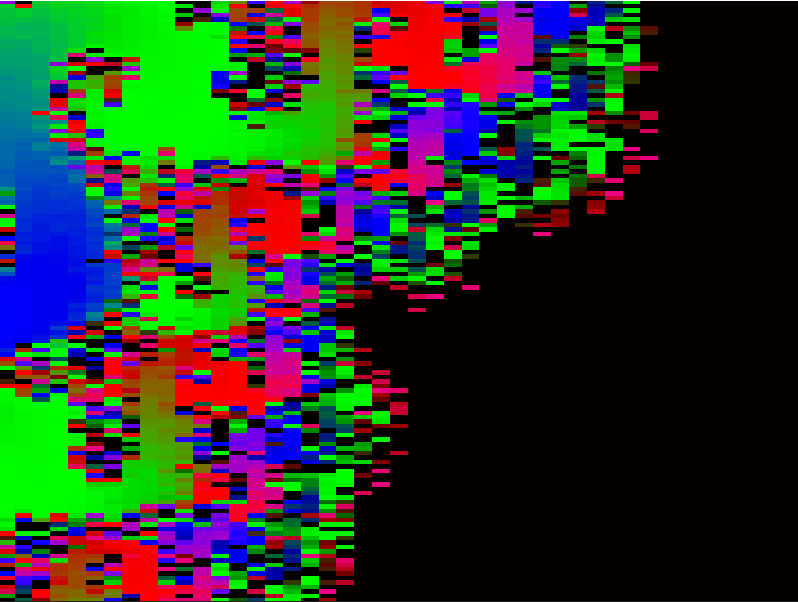} \\
\midrule
Zoom Factor: 1e-1 & Zoom Factor: 1e-4 & Zoom Factor: 1e-6 \\
\bottomrule
\end{tabular}
\caption{Visualization of the Mandelbrot Set at the Point Re=-0.7436450, Im=0.13182590 with Different Zoom Factors. The images display the fractal structure at zoom levels of 1e-1, 1e-4, and 1e-6, showcasing the intricate details at progressively finer scales. Precision concerns and pixelization problems can be seen when the zoom is 1e-6}
\label{tab:three_images}
\end{table}

\subsubsection{Rounding Issue}
Inaccuracies can occur due to the rounding issues. The first rounding may be toward a midpoint which then gets rounded again, moving it further from the closest correct value \cite{Boldo}. Consider the number 3.1415926 that is represented base-10. The higher precision will round to three decimal places while the lower precision will round to the nearest integer. The higher precision rounds 3.1415926 to 3.142. When this result is then rounded to a lower precision it becomes 3. When 3.1415926 is rounded directly to the nearest integer, omitting the intermediate step the answer is again 3 so in this case there is no inaccuracy. A slight modification of the situation can make double rounding significant. For instance, if the exact value was 3.6515926 then rounding first to the higher precision gives 3.652 and further rounding to the lower precision gives 4. Rounding directly from 3.6515926 to the nearest integer gives 4 also so differences need not appear in every case, yet are of vital significance in the vicinities of some number values \cite{lafage2020revisiting} \cite{tsarapkina2014exploring}.

\subsubsection{Limited Precision}
In general, limited precision refers to the extent of precision that can be attained in any computation or measurement. For computational and scientific purposes, it is extremely important to be valid in domains as diverse as numerical analysis and engineering since the accuracy and reliability of a result are determined by its precision. Single-precision floating-point values provide an approximate 7 decimal digits of precision \cite{10.5555/2071032}. This causes insufficient precision for more complex visualizations, and it may introduce substantial rounding errors with very large or small-scale numbers and lead to loss of details. 

\subsubsection{Range Limitations} 
This refers to being restricted to some values within a range that the computer system or a model of computation is capable of representing. These may be the largest or smallest values a number takes, either positive or negative. Fundamentally, they are directly proportional to the size of the data type used and are bound by limitations in the handling of very large or very small numbers. One issue with floats is that they have a restricted range, which can result in overflows or underflows. This could be manifested in graphical contexts as visual anomalies or inaccuracies during rendering \cite{8739150} \cite{https://doi.org/10.1002/pamm.202000079}.

\section{Comparative Analysis between Single-precision Floating Point and Double-precision Floating-point Implementations}

Most modern graphics and compute target applications rely on floating-point operation accuracy, which brings about dramatic performance and quality impacts. On the other side, single-precision versus double-precision floating-point implementations represent a compromise in both computational performance and accuracy for high-performance graphics rendering. This section describes a comparative study of both under the Vulkan API framework, pointing out each of their benefits and tradeoffs. In conjunction with the explanation, the diagram provided shows the Vulkan-based application — the full extent of which will initialize and manage GPU resources for rendering both 2D and 3D point datasets to cover in detail how the choice of precisions affects the resultant rendering and performance metrics.

\begin{figure} [H]
    \centering
    \includegraphics[width=1.0\linewidth]{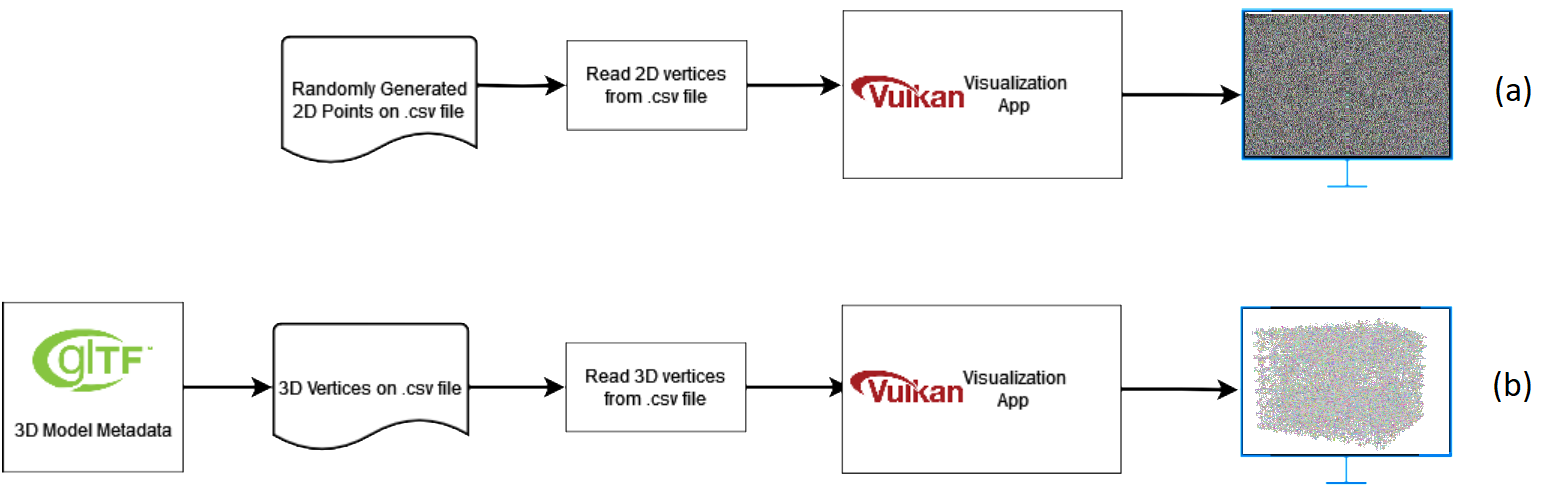}
    \caption{2D and 3D Point Data Visualization Process: a) 2D Point Data Visualization: The Vulkan-based visualization application reads a .csv file consisting of randomly generated 2D points: in this data set, every point is represented by the x and y coordinates. The CSV file is then read by the Vulkan-based visualization application, which visualizes it on the screen. (b) 3D Point Data Visualization: Downloaded model data in glTF/GLB format and further converted it into a .csv file, including the x, y, and z coordinates for each point. Feeding this .csv file into the Vulkan-based visualization application would draw the 3D points onto the screen. As soon as three-dimensional points can be visualized, more complex structures and models represented with data will easily be analyzed and understood.}
    \label{fig:enter-label}
\end{figure}

The first step is to create a Vulkan instance. A Vulkan instance is an instance that lets the application interact with the Vulkan API. Immediately following the creation of the Vulkan instance, a debug messenger is created for debugging during the development process. Following this, inter-functioning the windowing system with the surface creation is done. All core Vulkan API elements have been created at this stage, which would now allow an application to use the GPU.

\begin{figure*} 
    \centering
    \includegraphics[width=1.0\linewidth]{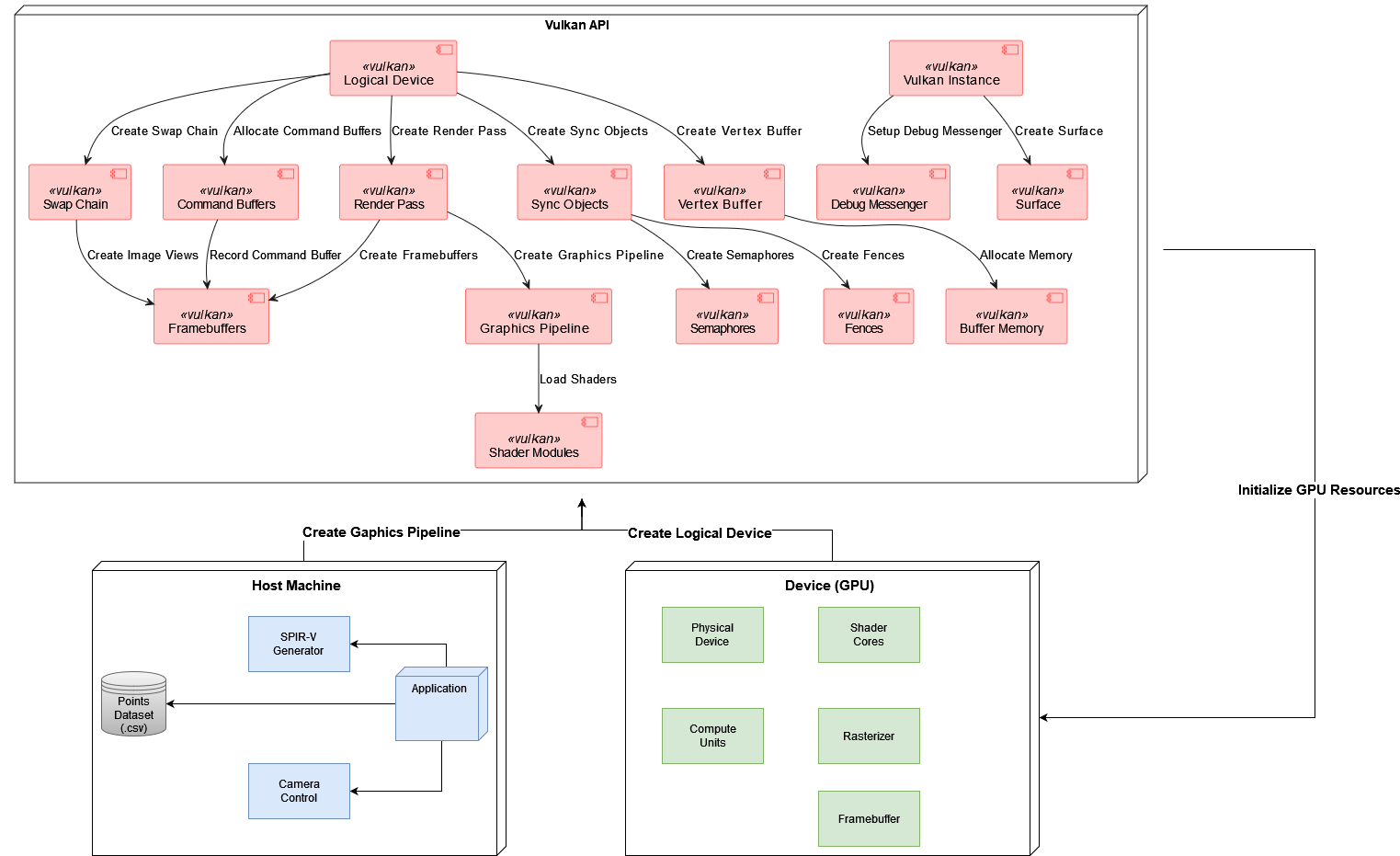}
    \caption{This diagram shows initializing and managing GPU resources using the Vulkan API to visualize 2D and 3D point datasets. Vulkan exposes a low-level, general-purpose graphics API that is conceived to offer direct control of the GPU resources to ensure both high performance and flexibility. It details all the processes, from the initialization of the GPU resources to the creation of the graphics pipeline. The graphics pipeline and the initialization of GPU resources are showcased. This diagram provides a comprehensive overview of the stages and interactions involved in setting up and utilizing Vulkan for rendering within the application.}
    \label{fig:enter-label}
\end{figure*}

Another core stage in managing the rendering of graphics using Vulkan is logical device setup. This logical device provides an interface with the GPU from the application. It enables a number of commands to be executed. At this stage, Command Buffers are allocated for the storage of rendering commands. Synchronization objects, like Semaphores and Fences, are created to control the synchronicity in the execution of commands. A Vertex Buffer is also allocated for vertex data. These parts ensure that all rendering processes go through smooth and concurrently.

This can do high-performance graphics operations with the Vulkan API by creating a Graphics Pipeline. It describes the process of rendering—a pipeline starting from processing vertices up to fragment shading. Shader Modules are loaded into the pipeline to handle certain parts of the rendering process. Rendering output is controlled by the arrangement of Render Passes and Framebuffers. Therefore with such structure, it is possible to execute efficiently complex graphics operations.

It involves all rendering commands, ance drawing 2D and 3D points, updating the Framebuffers, and, finally, memory allocation to manage one buffer—the so-called Buffer Memory—which contains vertex data and other related information regarding the rendering process. In this step, it will be finalized how an application is going to handle its rendering process to present the final output. Finally, memory is allocated to manage one buffer—the so-called Buffer Memory—which contains vertex data and other related information regarding the rendering process. Efficient memory management has been taken into consideration to ensure that.GUI resources are used effectively.

Control and data flow between Host Machine and GPU Host machine components: Points Dataset, SPIR-V Generator, Application, Camera Control. A Points Dataset, prepared, typically, in CSV format, contains 2D and 3D points. A SPIR-V Generator generates a low-level representation from an input high-level source representation, such as high-level shaders, for execution on a GPU. The Application is in charge of the entire rendering process and communicates with the Vulkan API. Camera Control deals with how the visualization is to be viewed.

The device components include the GPU, Physical Device, Shader Cores, Compute Units, Rasterizer, and Framebuffer. Each of these is one of the key elements in a rendering pipeline, and together they allow high performance in graphics-oriented operations.

Once the shaders are written in GLSL, they need to be compiled into SPIR‐V, which is the intermediate representation used by Vulkan. The compilation process ensures that the shaders are optimized and can be executed efficiently on the GPU. The compilation can be done using tools like `glslangValidator`. The compiled SPIR‐V shaders are then integrated into the Vulkan pipeline \cite{spirvtoolchain} \cite{whatisspirv}. The steps include creating shader modules, setting up the pipeline, and binding the necessary resources. Below is an example of how the shader modules are created and integrated:

The created shader module is then used in the graphics pipeline to execute the vertex and fragment shaders. By providing a detailed explanation of the shader code, its compilation, and integration into the Vulkan pipeline, this section offers a comprehensive understanding of how native double-precision floating-point operations are utilized in Vulkan applications.

\subsection{Emulated Double-precision Floating-point}
To compare the findings, case studies exemplifying emulated double-precision and native double-precision are presented. Development for emulated precision was based on David H. Bailey's approach in the DSFUN90 library \cite{baileydsfun90}.

\begin{algorithm}
\caption{Emulate Double Precision}
\begin{algorithmic}[1]
\Require $value \in \mathbb{R}$ \Comment{A double precision floating point number}
\Ensure $low, high \in \mathbb{R}$ \Comment{Two single precision floating point numbers representing the double precision input}
\Procedure{emulateDoublePrecision}{\textit{value}, \textit{low}, \textit{high}}
    \State $\textit{high} \gets (float) \textit{value}$
    \State $\textit{highDouble} \gets (double) \textit{high}$
    \State $\textit{low} \gets (float) (\textit{value} - \textit{highDouble})$
\EndProcedure
\State \Return $(\textit{low}, \textit{high})$
\end{algorithmic}
\end{algorithm}

This algorithm aims to store a double precision floating point number by dividing it into two single precision floating point numbers. First, the variable value is converted to a single precision floating point number and assigned to the variable high. This step is to obtain a lower precision representation of value. The high value is then converted back to a double precision number and assigned to the highDouble variable. This conversion is necessary for comparison with the original value. Finally, the difference is calculated by subtracting highDouble from the value, and this difference is assigned to the low variable as a single precision number. Thus, the variables high and low are stored as a two-part representation of the original double-precision value. This method is especially useful when precision is essential and memory savings are required.

\begin{lstlisting}[language=C, caption={Emulated Double-precision Floating-point Vertex Shader Code}, label={lst:art-owen}]
#version 450
layout(push_constant) uniform PushConstants {
    mat4 mvp;
} pushConstants;
layout(location = 0) in vec3 highPos;
layout(location = 1) in vec3 lowPos;
layout(location = 2) in vec3 highColor;
layout(location = 3) in vec3 lowColor;
layout(location = 0) out vec3 outHighColor;
layout(location = 1) out vec3 outLowColor;
void main() {
    outHighColor = highColor;
    outLowColor = lowColor;
    gl_Position = pc.mvp * vec4(highPos + lowPos, 1.0);
}
\end{lstlisting}

\begin{lstlisting}[language=C, caption={Emulated Double-precision Floating-point Fragment Shader Code}, label={lst:art-owen}]
#version 450
layout(location = 0) in vec3 outHighColor;
layout(location = 1) in vec3 outLowColor;
layout(location = 0) out vec4 fragColor;
void main() {
    vec3 fullColor = outHighColor + outLowColor;
    fragColor = vec4(fullColor, 1.0);
}
\end{lstlisting}
\subsection{Native Double-precision Floating-point}
Modern GPUs support double-precision floating-point natively, meaning that high-precision computation can be carried out without emulation. Vulkan API strongly supports native double-precision operations due to its shader language, GLSL (OpenGL Shading Language). In this work, double precision data types have been used, such as double and dvec2, performing vertex and fragment shaders in order to compute the exact values for the requested operations. Later, the shaders were compiled into SPIR‐V, which represents the Vulkan Intermediate Representation.

\begin{lstlisting}[language=C, caption={Native Double-precision Floating-point Vertex Shader Code}, label={lst:art-owen}]
#version 450
#extension GL_ARB_gpu_shader_fp64 : enable
layout(push_constant) uniform PushConstants {
    dmat4 mvp;
} pushConstants;
layout(location = 0) in dvec3 pos;
layout(location = 1) in dvec3 color;
layout(location = 0) out flat dvec3 fragColor;
void main() {
    gl_Position = vec4(pc.mvp * dvec4(pos, 1.0));
    fragColor = color;
}
\end{lstlisting}
This is a vertex shader written in GLSL, version 4.5.0, with the GL\_ARB\_gpu\_shader\_fp64 extension. The program uses a push constant block called PushConstants, including a member of type dmat4 for an MVP matrix. It also has an input of type dvec3 for position and color, and finally defines a variable named fragColor of data type flat dvec3 to create its output. In the main function, declare a dvec4 vector with a pos data and 1.0. Multiply the vector by the MVP matrix and assign it to gl\_Position. Finally, assign the input color to the fragColor variable.

\section{Experimental Results}

\begin{figure} [H]
    \centering
    \includegraphics[width=1.0\linewidth]{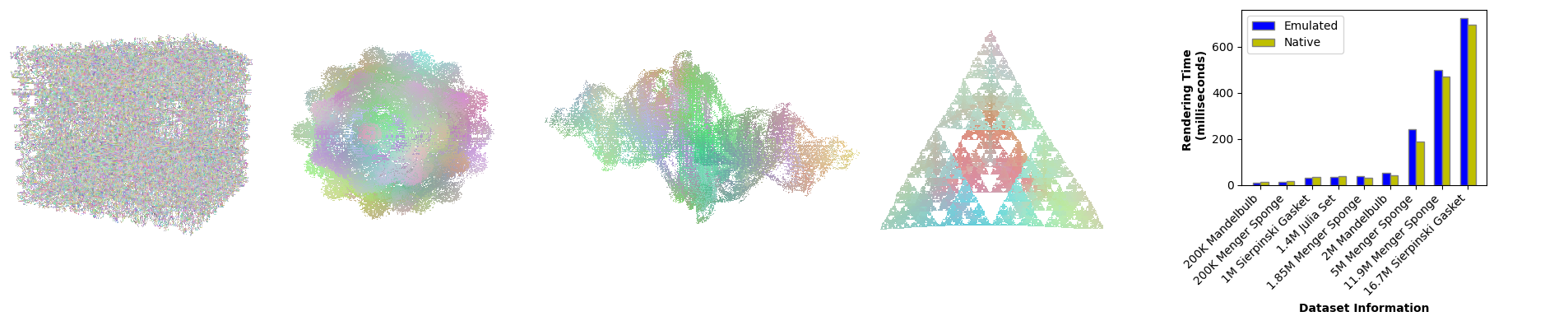}
    \caption{A few of the visualized objects from 3D datasets with double-precision floating-points are included, and the column chart on the right-hand side shows a comparison of rendering times in milliseconds for emulated double-precision and native double-precision.}
    \label{fig:enter-label}
\end{figure}

The proposed method has been implemented using Vulkan API, C++ 20, and GLSL, using only vertex shaders and fragment shaders compiled into SPIR-V code. To be used for testing, 2D point data consisting of 10,000, 100,000, 1,000,000, and 10,000,000 (x, y) coordinates were created randomly and uniquely in the range (-1.0 and 1.0) and then saved in .csv files. 3D point data consisting of 200,000 to 16,700,000 (x,y,z) coordinates were acquired 3D fractal models that created with the open sources libraries. The algorithms and open-source libraries used for dataset production are explained in Appendix A. The same datasets were used for both emulated double-precision experiments and native double-precision experiments. The experiments were conducted using NVIDIA RTX 3090 GPU, Intel(R) Core(TM) i7-6850K @3.60GHz CPU, and 44 GB DDR4 RAM hardware components. The framerate was recorded using the RenderDoc \cite{karlsson2019renderdoc} v1.33.

\begin{figure} [H]
    \centering
    \includegraphics[width=1.0\linewidth]{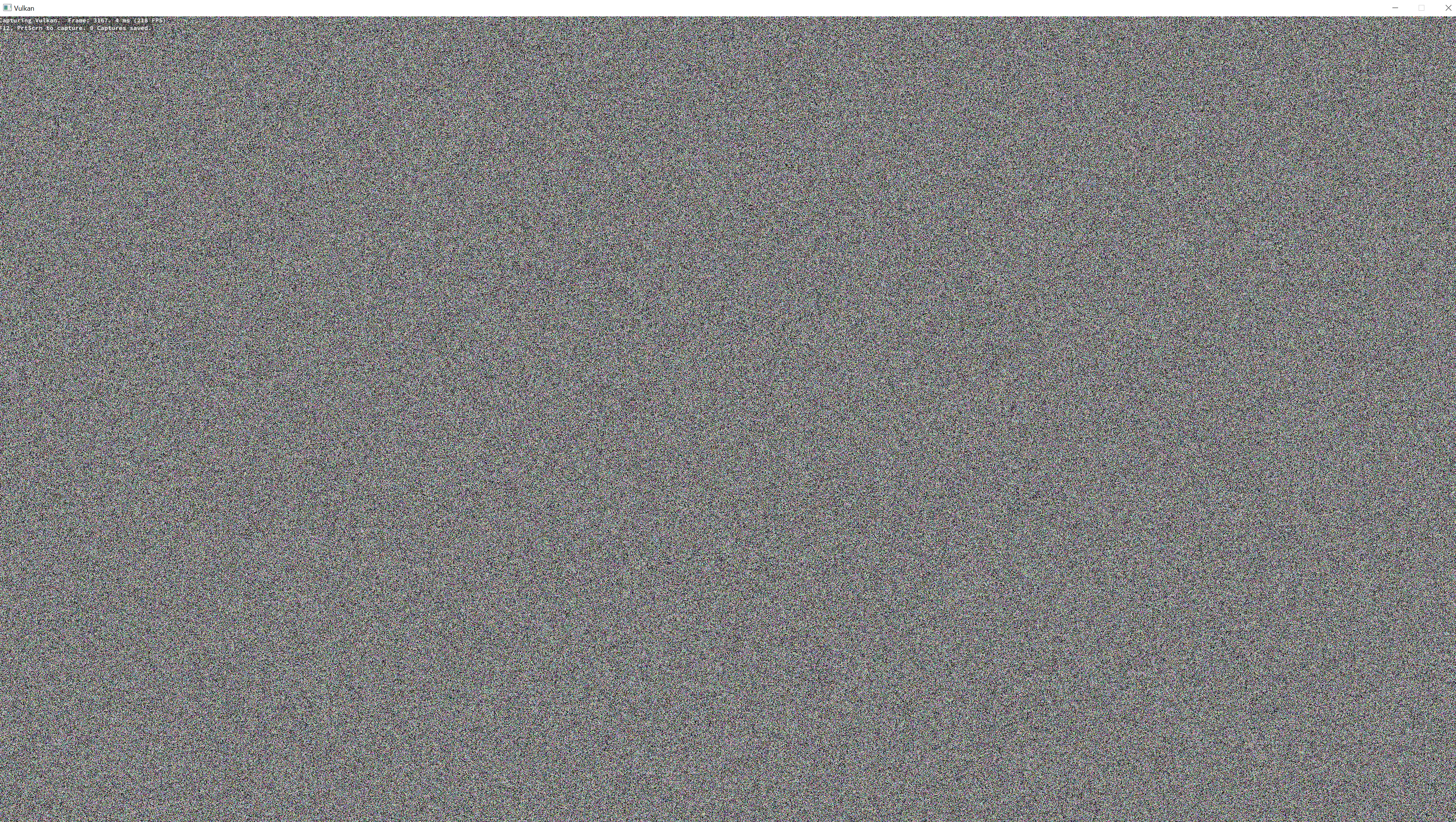}
    \caption{This figure is a screenshot of the application of an example 2D dataset. This dataset has 10 million 2D point data double-precision and the framerate value is calculated as 218 fps.}
    \label{fig:enter-label}
\end{figure}

\begin{figure} [H]
    \centering
    \includegraphics[width=0.7\linewidth]{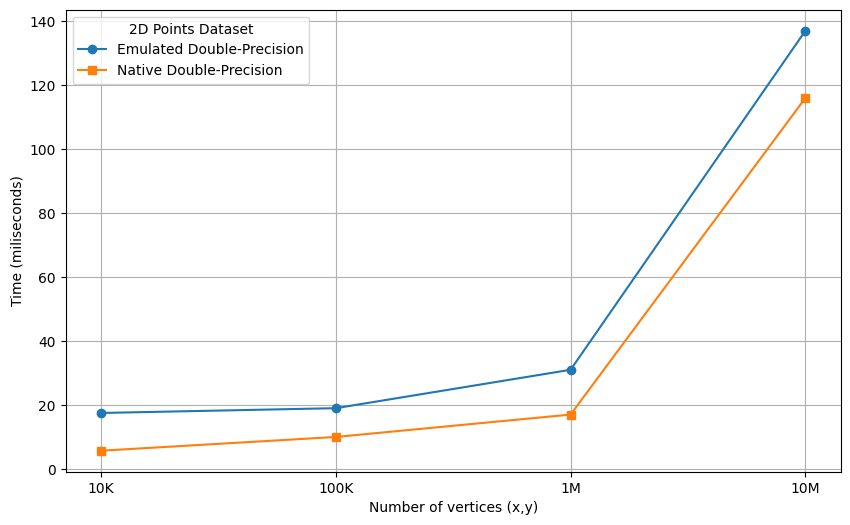}
    \caption{The graph shows the performance of both approaches in time (seconds) for different vertex numbers (10K, 100K, 1M, and 10M). The results reveal that native double-precision calculations are faster overall. The emulated double precision shows a significant performance degradation, especially at high vertex counts (10M). This finding indicates that local double-precision calculations are a more efficient option for calculations requiring high precision. While the emulation performs reasonably well at low vertex counts, local calculations are significantly faster on larger-scale datasets. This is critical for developers who want to perform high-precision and high-performance graphics operations using the Vulkan API.}
    \label{fig:enter-label}
\end{figure}

\begin{figure} [H]
    \centering
    \includegraphics[width=0.7\linewidth]{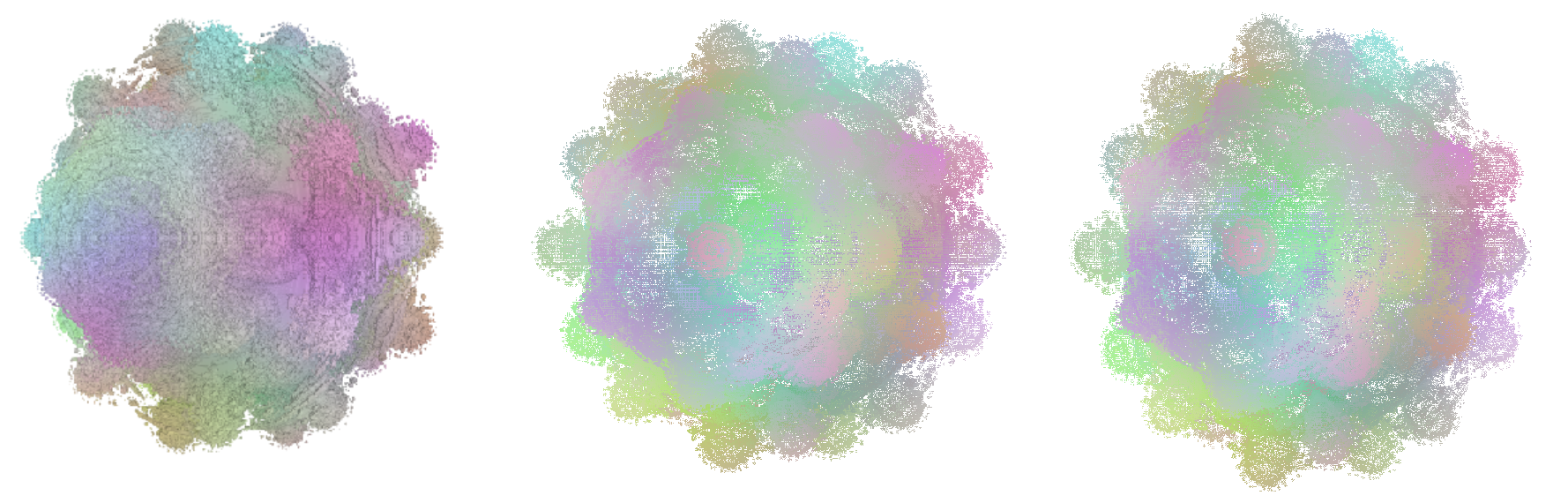}
    \caption{The images prove that three renderings are realized from the same dataset. The left shape represents a triangulated Mandelbulb mesh in GLB/GLTF format; the middle image offers a rendering of those vertices in native double-precision; and the rightmost image, the output of the emulated double-precision implementation run against the same mesh vertices. Comparisons of the performance of these two were done, and because the dataset being fed is the same, the generated renderings are the same. This study compares rendering performance and accuracy for both the native double precision and the emulated double precision methods for rendering.}
    \label{fig:enter-label}
\end{figure}

Experiments demonstrate the advantages of using native double-precision arithmetic within Vulkan. Performance measurements indicate that rendering time for a large dataset improves significantly. The results for native double precision are summarized in the tables below.

\begin{table} [H]
    \centering
    \begin{tabular}{|p{5.5cm}|p{2cm}|p{2cm}|} \hline 
        Dataset Information & Rendering Time (milliseconds) & Framerate (fps)\\ \hline 
        200K vertices of 3D Mandelbulb &  9.71 & 729 \\ \hline 
        200K vertices of 3D Menger Sponge & 12.9 & 703 \\ \hline 
        1M vertices of 3D Sierpinski Gasket & 32.19 & 745\\ \hline
        1.4M vertices of 3D Julia Set & 35.26 & 717\\ \hline
        1.85M vertices of 3D Menger Sponge & 36.69 & 612\\ \hline 
        2M vertices of 3D Mandelbulb &  53.19 & 687 \\ \hline         
        5M vertices of 3D Menger Sponge & 239.73 & 698 \\ \hline
        11.9M vertices of 3D Menger Sponge & 499.12 & 468\\ \hline
        16.7M vertices of 3D Sierpinski Gasket & 722.89 & 115\\ \hline
    \end{tabular}
    \caption{Performance Results for Emulated Double-precision Floating-point Implementations.}
    \label{tab:my_label}
\end{table}

The data in Table 2 indicates that, in general, the emulated double-precision floating point calculations are worse when compared to native calculations. Generally, the render times are longer and frame rates are lower. For example, the render time for a 3D Menger Sponge with 11.9 million vertices was up to 499.12 milliseconds. The highest frame rate observed was 729 fps for 3D Mandelbulb with 200K vertex.
Native Double Precision Floating Point Calculations

\begin{table}[H]
    \centering
    \begin{tabular}{|p{5.5cm}|p{2cm}|p{2cm}|} \hline 
        Dataset Information & Rendering Time (milliseconds) & Framerate (fps)\\ \hline 
        200K vertices of 3D Mandelbulb &  11.87 & 842 \\ \hline 
        200K vertices of 3D Menger Sponge & 14.88 & 804 \\ \hline 
        1M vertices of 3D Sierpinski Gasket & 35.26 & 789\\ \hline
        1.4M vertices of 3D Julia Set & 39.26 & 802\\ \hline
        1.85M vertices of 3D Menger Sponge & 31.12.16 & 763\\ \hline 
        2M vertices of 3D Mandelbulb &  42.68 & 854 \\ \hline         
        5M vertices of 3D Menger Sponge & 189.25 & 802 \\ \hline
        11.9M vertices of 3D Menger Sponge & 471.29 & 593\\ \hline
        16.7M vertices of 3D Sierpinski Gasket & 695.76 & 330\\ \hline
    \end{tabular}
    \caption{Performance Results for Native Double-precision Floating-point Implementations.}
    \label{tab:my_label}
\end{table}

Data in Table 3 clearly indicates that natively provided double precision excels in performance compared to emulated calculations. In most cases, render times are shorter, along with frame rates that become higher. For instance, 3D Menger Sponge containing 11.9 million vertices dropped the render time to as short as 471.29 milliseconds, beating emulated calculations. The highest frame rate among the datasets was achieved with 3D Mandelbulb 200K vertex at 854 fps.

Generally, the performance of natively conducted double-precision computations is better on a large dataset. Though there is droppings performance in emulated computations, the native ones are much more stable. This supports that the local computations are more appropriate when working on large data sets.

\section{Limitations}
The research demonstrates significant improvements in graphic visualization under the Vulkan API using double-precision floating-point data. However, several limitations should be realized. The native double-precision implementation is highly dependent upon modern GPU availability and capability, thereby limiting such an approach to older/less powerful hardware. Although the proposed method has much better scalability with large data sets than emulated double-precision, challenges to efficiently processing extremely large datasets may exist that decrease performance gains with dataset size. Also, the experimental results were obtained on specific hardware and software configurations and may not generalize to other systems; additional benchmarking on a diversity of configurations is thus required. Still, however illuminating the controlled setting with 2D and 3D point datasets was for this method, further testing on real-world data is required to confirm its applicability and performance in practical scenarios since different kinds of data will raise unique issues and possibly differing performance characteristics.

\section{Conclusion and Future Works}
This study was conducted to compare the performance and accuracy of Vulkan API, explicitly focusing on mutable and double-precision data types. The results show enormous implications for data type choices, more so concerning calculation speed and processing time. Double-precision solutions have been executed on the GPU at the moment using double-float and double-double techniques. Test applications consist of 2D points and 3D points; each of them contains double-precision vertex coordinates.

Necessary information is given in the paper hand, and experiments on using double-precision directly have been performed in the case of supporting GPU hardware. The work straight followed a method of double precision ordered by the Khronos Group in OpenGL Shading Language specification 4.5 and Vulkan specification 1.3. This is a different approach compared to traditional emulated precision methods, with no extra processing required. Although this method requires both advanced hardware and software, this is an example of the studies in scientific visualization where precision and performance are desirable together. In this work, the fundamental layer of the visualization of 2D points by uniquely generating random x and y values for these points has been addressed, and 3D points by generating from 3D mesh models with double precision x,y, and z values.

In the future, several experiments can be performed—the ones using actual data to make the application more applicable and accurate. These datasets can be used in tests that verify the visualizing methods. In the case of successful implementation of the most straightforward building block of graphical visualization, that is, point rendering, additional visualization components can be integrated to compile a full-featured visualization library. This will provide the ability to create more complex and informative visualizations, therefore increasing the application's functionality beyond simple point rendering.

\appendix
\section{Datasets}
To provide a more challenging and realistic evaluation, 3D point data were prepared using fractal algorithms, including Mandelbulb, Menger Sponge, Sierpinski Gasket and Julia fractals. The use of fractal models provides a complex and intricate dataset to test the rendering capabilities and performance of the proposed method in three-dimensional space. The diverse point counts ensure that the method's efficiency and effectiveness can be evaluated under different levels of complexity. 

Advanced libraries and algorithms were employed in this research:
\begin{itemize}
  \item NumPy (\url{https://numpy.org/}) and Pandas (\url{https://pandas.pydata.org/}): Used for data handling and manipulation.
  \item Skimage (\url{https://scikit-image.org/}): Applied for surface extraction and mesh generation.
  \item Pygltflib (\url{https://pypi.org/project/pygltflib/}): Utilized to store outputs in GLTF2 format.
  \item Numba (\url{https://numba.pydata.org/}): Provided computation acceleration with jit and prange functions.
  \item Open3D (\url{https://www.open3d.org/}): Used for 3D data visualization.
  \item Noise (\url{https://pypi.org/project/noise/}) library's pnoise3 function: Generated noise data for fractal colors.
\end{itemize}

All this has been applied to the fine development of such important complex fractal structures as Mandelbulb, Menger Sponge, Julia, and Sierpinski Gasket, with double precision x, y, z local coordinates, and RGB color values. The meshes that are generated manifest high visual quality and generally accurate details for scientific analyses. All datasets and their source codes can be accessible on this repository: \url{https://github.com/NeziheSozen/3d-fractal-generators}

The Mandelbulb is a three-dimensional, mathematical object of fractal nature \cite{white2009unravelling} \cite{barrallo2010expanding} \cite{knill2023mandelbulb}. This paper uses the Daniel White and Paul Nylander approach using spherical coordinates. The following algorithm was used to generate the 3D Mandelbulb datasets:

\begin{algorithm} [H]
\caption{Mandelbulb 3D Fractal Creation}
\begin{algorithmic}
\State Input: $max\_iterations$, $bailout$, $power$, $resolution$, $x_{\min}$, $x_{\max}$, $y_{\min}$, $y_{\max}$, $z_{\min}$, $z_{\max}$
\State Output: Mandelbulb 3D fractal

\For{$x \gets x_{\min}$ to $x_{\max}$ step $resolution$}
    \For{$y \gets y_{\min}$ to $y_{\max}$ step $resolution$}
        \For{$z \gets z_{\min}$ to $z_{\max}$ step $resolution$}
            \State $z_x \gets x$, $z_y \gets y$, $z_z \gets z$
            \State $iteration \gets 0$
            \While{$iteration < max\_iterations$  and $(z_x^2 + z_y^2 + z_z^2) < bailout$}
                \State $r \gets \sqrt{z_x^2 + z_y^2 + z_z^2}$
                \State $\theta \gets \arctan2(\sqrt{z_x^2 + z_y^2}, z_z)$
                \State $\phi \gets \arctan2(z_y, z_x)$
                \State $new\_r \gets r^\text{power}$
                \State $new\_theta \gets \theta \cdot \text{power}$
                \State $new\_phi \gets \phi \cdot \text{power}$
                \State $z_x \gets new\_r \cdot \sin(new\_theta) \cdot \cos(new\_phi) + x$
                \State $z_y \gets new\_r \cdot \sin(new\_theta) \cdot \sin(new\_phi) + y$
                \State $z_z \gets new\_r \cdot \cos(new\_theta) + z$
                \State $iteration \gets iteration + 1$
            \EndWhile
            \If{$iteration = max\_iterations$}
                \State Point $(x, y, z)$ is part of the fractal.
            \Else
                \State Point $(x, y, z)$ is not part of the fractal.
            \EndIf
        \EndFor
    \EndFor
\EndFor
\end{algorithmic}
\end{algorithm}

This algorithm creates points in a 3-dimensional space by calculating the Julia set with a given quaternion and parameters \cite{norton1989julia} \cite{hart1990interactive}:

\begin{algorithm}[H]
\caption{Quaternion Based Julia Set Algorithm}
\begin{algorithmic}
\Function{JuliaSetQuaternion}{$c, \text{max\_iter}, \text{threshold}, \text{resolution}$}
    \For{$x \gets -\text{resolution}$ \textbf{to} $\text{resolution}$}
        \For{$y \gets -\text{resolution}$ \textbf{to} $\text{resolution}$}
            \For{$z \gets -\text{resolution}$ \textbf{to} $\text{resolution}$}
                \State $q \gets \text{quaternion}(x, y, z, 0)$
                \State $n \gets 0$
                \While{$n < \text{max\_iter}$ \textbf{and} $\text{norm}(q) < \text{threshold}$}
                    \State $q \gets q^2 + c$
                    \State $n \gets n + 1$
                \EndWhile
                \If{$\text{norm}(q) \geq \text{threshold}$}
                    \State \text{plot\_point}$(x, y, z, n)$
                \EndIf
            \EndFor
        \EndFor
    \EndFor
\EndFunction
\end{algorithmic}
\end{algorithm}

The Menger sponge is a three-dimensional fractal geometric shape defined by Karl Menger. A structure such as this one is developed by removing smaller cubes from the center and each face of an initial cube, therefore making a structure of almost no volume and infinite surface area through its infinite iterations. \cite{edgar2019classics} \cite{peitgen2004chaos}

\begin{algorithm} [H]
\caption{Menger Sponge Creation}
\begin{algorithmic}
\State Input: $max\_iterations$, $cube\_size$
\State Output: Menger Sponge

\Function{MengerSponge}{$x, y, z, size, iteration$}
    \If{$iteration = max\_iterations$}
        \State Draw cube at $(x, y, z)$ with size $size$
    \Else
        \State $new\_size \gets size / 3$
        \For{$i, j, k \in \{0, 1, 2\}$}
            \If{not $(i = 1 \land j = 1) \lor (i = 1 \land k = 1) \lor (j = 1 \land k = 1)$}
                \State \Call{MengerSponge}{$x + i \cdot new\_size, y + j \cdot new\_size, z + k \cdot new\_size, new\_size, iteration + 1$}
            \EndIf
        \EndFor
    \EndIf
\EndFunction

\State \Call{MengerSponge}{$0, 0, 0, cube\_size, 0$}
\end{algorithmic}
\end{algorithm}

The Sierpinski Gasket also referred to as the Sierpinski Triangle, is named after by the name of Wacław Sierpiński, who described this fractal \cite{sierpinski2020general}. The creation of this fractal involves recursively cutting an equilateral triangle into three smaller equilateral triangles and leaving the central triangle of each such division empty. This is done as many times as possible. Thus, it goes on to yield an extremely elaborate and self-replicating pattern \cite{Whitrow2008}. The below algorithm shows the procedures to create The Sierpinski Tetrahedron:

\begin{algorithm} [H]
\caption{Sierpinski Gasket Tetrahedron}
\begin{algorithmic}[1]
\Procedure{GenerateSierpinskiTetrahedron}{}
    \State \textbf{Input:} Number of iterations $n$
    \State \textbf{Output:} Sierpinski gasket tetrahedron

    \State Initialize tetrahedron with vertices $A, B, C, D$
    \State \textbf{for} $i = 1$ \textbf{to} $n$ \textbf{do}
        \State Divide each face of the tetrahedron into 3 equal parts
        \State Calculate the midpoints of each face:
        \State $M_{AB} = \frac{A + B}{2}$
        \State $M_{BC} = \frac{B + C}{2}$
        \State $M_{CA} = \frac{C + A}{2}$
        \State Create new tetrahedrons using these midpoints
        \State Replace the original tetrahedron with the new tetrahedrons
    \State \textbf{end for}
\EndProcedure
\end{algorithmic}
\end{algorithm}

\bibliographystyle{unsrt}  
\bibliography{references}

\end{document}